\documentclass[journal]{IEEEtran}
\usepackage{times}
\usepackage{lipsum}
\usepackage{mwe}
\usepackage{fancybox}
\usepackage{multicol}
\usepackage{color}
\usepackage{graphicx}
\usepackage{epsfig}
\usepackage{graphicx,latexsym}
\usepackage{amssymb,amsthm,amsmath}
\usepackage{array}
\usepackage{xy}
\usepackage{amsthm}
\usepackage{mathtools}
\usepackage{epstopdf}
\usepackage{algorithm}
\usepackage{algpseudocode}
\usepackage{cite}
\usepackage{ amssymb }
\usepackage[table,xcdraw]{xcolor}
\usepackage{float}
\usepackage{multirow}

\usepackage{hhline}
\usepackage{url}
\usepackage{booktabs}
\usepackage{amsmath,algorithm,tabularx}
\usepackage{subcaption}
\usepackage{dblfloatfix}

\newcolumntype{C}[1]{>{\centering\arraybackslash$}p{#1}<{$}}
\makeatletter

\newcommand{\Rmnum}[1]{\expandafter\@slowromancap\romannumeral #1@}

\newcommand{\multiline}[1]{%
  \begin{tabularx}{\dimexpr\linewidth-\ALG@thistlm}[t]{@{}X@{}}
    #1
  \end{tabularx}
}
\makeatother
\begin{document}
\title{{Slicing-Based AI Service Provisioning on Network~Edge}}
\author{Mushu~Li,~\IEEEmembership{Student~Member,~IEEE,}
        Jie~Gao,~\IEEEmembership{Member,~IEEE,}
        Conghao~Zhou,~\IEEEmembership{Student~Member,~IEEE,}
        Xuemin~(Sherman)~Shen,~\IEEEmembership{Fellow,~IEEE,} 
        and~Weihua~Zhuang,~\IEEEmembership{Fellow,~IEEE} 
        \thanks{This work has been submitted to the IEEE for possible publication. Copyright may be transferred without notice, after which this version may no longer be accessible.
        
        Mushu Li, Conghao Zhou, Xuemin (Sherman) Shen, and Weihua Zhuang are with the Department of Electrical and Computer Engineering, University of Waterloo, Waterloo, ON, Canada, N2L 3G1 (email: \{m475li, c89zhou, sshen, wzhuang\}@uwaterloo.ca). 
        
Jie Gao is with the Department of Electrical and Computer Engineering, Marquette University, Milwaukee, WI, USA, 53233 (e-mail: j.gao@marquette.edu).
}
}%

\maketitle
\thispagestyle{empty}
\begin{abstract}
Edge intelligence leverages computing resources on network edge to provide artificial intelligence (AI) services close to network users. As it enables fast inference and distributed learning, edge intelligence is envisioned to be an important component of 6G networks.
In this article, we investigate AI service provisioning for supporting edge intelligence. 
First, we present the features and requirements of AI services. Then, we introduce AI service data management, and customize network slicing for AI services. Specifically, we propose a novel resource pooling method to jointly manage service data and network resources for AI services.
A trace-driven case study demonstrates the effectiveness of the proposed resource pooling method. Through this study, we illustrate the necessity, challenge, and potential of AI service provisioning on network edge.

\end{abstract}

\section{Introduction}

The sixth-generation (6G) networks are envisioned to support many emerging use cases, such as extended reality, remote healthcare, and autonomous systems~\cite{OJVT,Saad}. Compared with services supported by the fifth-generation (5G) networks, services in the 6G era will be even more diversified, potentially blurring the boundaries among enhanced mobile broadband (eMBB), massive machine-type communications (mMTC), and ultra-reliable and low-latency communications (URLLC). Such services will demand highly intelligent and flexible networks, driving a confluence of advanced networking and artificial intelligence (AI) technologies.

AI can play an essential role in network management, e.g., resource management~\cite{Wu,TCCN} and protocol design~\cite{MAC}. Meanwhile, AI can also be a new type of service supported by networks. Examples of AI services in future networks include language processing, video surveillance, and autonomous driving. Since AI services will need to gather or generate a vast amount of data, edge intelligence has attracted extensive interest as it moves AI closer to user devices (UDs) and alleviates data traffic load in the core network. Empowered by distributed learning techniques, edge intelligence leverages the communication, computing, and storage resources at each edge node, i.e., a base station (BS) and other access points (APs), to execute data processing and inference tasks.

Typically, an AI service involves two phases, i.e., inference and model training.
Different from conventional services, AI services largely depend on the data generated by UDs, and such dependence exists in both phases. For example, image recognition services depend on images and corresponding labels uploaded from UDs. As a result, the availability and quantity of data from UDs determine the effectiveness of an AI service, such as inference accuracy and learning rate. Inference accuracy may increase when more data is available at an edge node. The available data for an AI service at an edge node consists of two parts, i.e., data collected from UDs by this edge node and data from other edge nodes. While sharing data among all edge nodes increases the amount of available data and potentially improves the performance of AI services, it requires significant communication and computing resources. Specifically, each edge node needs excessive computing resource for data processing and communication resource for exchanging data with other edge nodes. Considering that the amount of data collected by each edge node can be very different, a viable alternative to sharing all data is to migrate a portion of data from edge nodes that have collected sufficient data to those that need more. Achieving this requires scalable and on-demand network resource management, especially considering that AI services need to co-exist and share resources with conventional services. While a few existing works, such as~\cite{Yang} and~\cite{Ren}, have studied the relation between the performance of AI services and network resource allocation, the topic needs further investigation.

As a major innovation in 5G technology, network slicing can support a multitude of network services with diverse service requirements by creating and maintaining logically isolated virtual networks, i.e., slices, for different services~\cite{OJVT}. Network slicing has a potential to support AI services in future networks. However, due to the unique features and requirements of AI services, a slicing-based network should not treat an AI service in the same way as any conventional services. The reason is two-fold. First, the location of physical resources can have a non-negligible impact on the performance of AI services, while in network slicing, resources are allocated from a virtual resource pool and the location of resources is hidden through abstraction. Second, AI services have unique performance metrics, such as accuracy, that require coordination of data available to the edge nodes, while network slicing considers conventional performance metrics, such as throughput and delay.


\begin{figure*}[t]  
  \centering  
  \includegraphics[width=185mm]{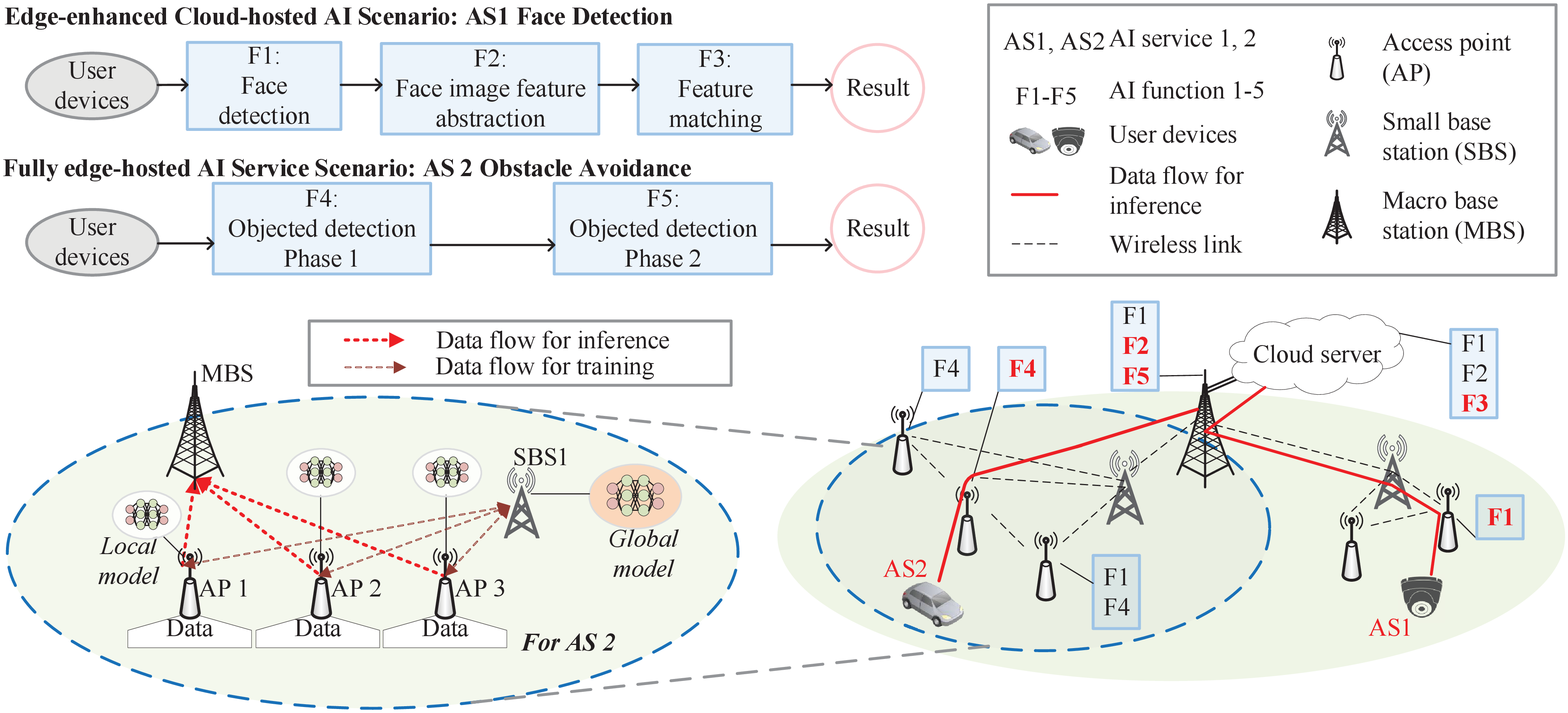}\\
  \caption{Service management for AI services on network edge.}  
  \label{fig:ais} 
\end{figure*}

In the following, we investigate AI service provisioning on network edge and extend network slicing to support AI services. Specifically, we propose a resource pooling method, which customizes resource virtualization for each AI service by considering the location of physical resources and enabling effectual data migration among edge nodes. The proposed resource pooling method addresses the aforementioned challenges in the existing network slicing framework.  
Furthermore, we provide a case study to demonstrate the effectiveness of the resource pooling method for AI services. 

\section{AI Services and Requirements}\label{sec:AISR}

\subsection{AI Services at Network Edge}

Similar to conventional services, an AI service is enabled by a chain of functions. The difference is that one or more functions are based on AI models, such as deep neural networks (DNNs) and \textit{k}-nearest neighbors algorithms. We refer to such functions as AI functions. In existing networks, AI functions are mostly deployed in a cloud server, while edge nodes simply forward the data of UDs to the cloud server. The disadvantage of such cloud-centric AI service provisioning is heavy data traffic load on the core network. To address this issue, some AI functions can be deployed at the network edge to be close to UDs. Specifically, edge nodes can play an active role to support AI services in the following scenarios:
\begin{itemize}
\item Edge-assisted cloud-hosted AI scenario:  A small portion of AI functions, such as data preprocessing and aggregation, are deployed at the network edge, while the rest of AI functions are executed at the cloud server;
\item Cloud-assisted edge-hosted AI scenario: All AI functions are placed at the network edge for inference, and the cloud server assists the network edge in training the AI models used by AI functions. The cloud server coordinates the exchange of information, such as DNN parameters or gradients, among edge nodes. An example is shown as AI service 1 in Fig.~\ref{fig:ais};
\item Fully edge-hosted AI scenario: All AI functions are deployed at the network edge, and the edge nodes exchange information with each other for training AI models. An example is shown as AI service 2 in Fig.~\ref{fig:ais}.
\end{itemize}
A comparison of the above three scenarios is summarized in Table~\ref{table}. 
\begin{table*}[t]
\centering{
\caption{Three scenarios of AI services in edge intelligence}\label{table}
\begin{tabular}{|l|l|l|l|}
\hline
\textit{}                                      & \textbf{Edge-assisted cloud-hosted AI scenario}                                                                                               & \textbf{Cloud-assisted edge-hosted AI scenario}                                                                                                                     & \textbf{Fully edge-hosted AI scenario}                                                                                                          \\\hline 
\textbf{Use cases}                                                          & Image and voice recognition                                                                               & Automated driving, mobile VR                                                                                           & Business informatics, smart city                                                                     \\\hline 
\textbf{\begin{tabular}[c]{@{}l@{}} Key resource in\\   demand\end{tabular}} & Communication                                                                                & Computing                                                                                                              & Computing                                                                              \\ \hline
\textbf{Role of edge}                                                      & Data preprocessing and aggregation                                                                        & Inference                                                                                                                  & Inference and model training                                                               \\ \hline
\textbf{\begin{tabular}[c]{@{}l@{}} Requirements and \\  features  of service \end{tabular}}                                                   & \begin{tabular}[c]{@{}l@{}}Large  data size that requires database in\\  the cloud server\end{tabular} & \begin{tabular}[c]{@{}l@{}}Stringent service requirement that requires\\   real-time training and fast inference \end{tabular} & \begin{tabular}[c]{@{}l@{}}Demand for fast inference and privacy\\-preserving measures \end{tabular} \\ \hline
\textbf{Learning methods}                                                   & Centralized and offline learning                                                                  & Federated learning, splitting learning, etc.                                                                                         & Transfer learning, gossip learning, etc.                                                             \\ \hline
\end{tabular}}
\end{table*}
\subsection{Key Performance Indicators}

Since AI services can be viewed as a special type of computing services, conventional performance indicators such as latency and energy efficiency apply to AI services. In addition,  the following new performance indicators are necessary for evaluating the performance of AI services:
\begin{itemize}
\item Accuracy~\cite{Yang, Ren}, which measures the difference between inference results derived by an AI service and the real values; 
\item Learning speed~\cite{li2019convergence,Ren}, which measures how fast an AI model can be fully trained. For example, in the case of DNN, the learning speed is the convergence rate of the loss function during the training process.
\end{itemize}
Moreover, other performance indicators,  such as running time~\cite{Qingqing} and memory shrinks~\cite{Vanhoucke}, can also be applied for evaluating certain AI models used by an AI service.

\subsection{Features of AI Services}
\label{sec:AIfeature}
In general, an AI service consists of two phases, i.e., inference and model training. 
In inference, edge nodes process data from UDs and deliver computing results to UDs or other edge nodes, which is similar to conventional computing services.
For model training, the data available to an edge node includes the data collected from UDs and the data migrated from other edge nodes. 
Each edge node utilizes its available data to train the AI  models used by AI functions and exchange training parameters with other nodes to improve the effectiveness of training. For example, in federated learning, edge nodes train their local AI models, upload the parameters of local models to a centralized node, and obtain the parameters of a global model from the centralized node periodically.
 



For both inference an training, data flows from UDs to edge nodes as well as among edge nodes are necessary. In the inference phase, the way that data flows among edge nodes affects the performance, e.g., inference delay, of AI functions. 
In the training phase, the way that data flows among edge nodes affects the performance of AI functions from the following three aspects. 
First, the migration of data among edge nodes determines how model training is performed. 
We define a term, learning structure, to specify which edge nodes train the AI model and how they migrate data with each other in the network.
If data from UDs is migrated to fewer edge nodes for training, the learning structure is more centralized, and the benefit is a higher learning speed and inference accuracy.
Second, the migration of data among edge nodes balances the available data at the edge nodes and alleviates data bias. This can further improve inference accuracy~\cite{Wang, ntoutsi2020bias}  and speed up loss function value convergence for distributed learning~\cite{li2019convergence}.
Third, in addition to migrating data collected from UDs, the training parameter exchange among edge nodes affects the learning speeds of AI models. For example, frequent model aggregation in federated learning leads to fast convergence at the cost of high data traffic volume among edge nodes.


\subsection{Service Data Management}
\label{sec:AIServicemanagement}

Given the potential impacts of data flow on the performance of AI services, service data management is required for AI services, which includes AI function placement, AI model parameter selection, and AI service operation.

\textbf{AI function placement:} 
The functions of an AI service can be executed at edge nodes. An AI function placement policy determines which edge nodes are selected to host AI functions. An example of AI function placement is illustrated in Fig.~\ref{fig:ais}. In the inference phase, APs~1 to~3 and the Marco BS (MBS) provide inference for the UDs for AI service AS~2. In the model training phase, the APs upload and download AI models to/from Small BS (SBS)~2 to train the AI model in function F4. Data flows through AI functions deployed at edge nodes for both inference and model training. By placing AI functions at edge nodes, data flow can be initialized, and the learning structure for AI services can be defined.


\textbf{AI model parameter selection:} 
AI model parameters include learning rate for DNNs and model aggregation frequency in federated learning.
Based on AI function placement,  edge nodes train the AI models adopted by AI functions through exchanging the AI model parameters with each other. The parameters of AI models affect the AI service performance, e.g., accuracy and learning speed. Moreover, they specify the parameter exchange frequency and the amount of data for parameter exchange among edge nodes over time. 


\textbf{AI service operation:} AI service operation is responsible for scheduling data flow in real time according to network conditions, such as channel conditions and instantaneous computing latency of edge nodes, given AI function placement and AI model parameters.
For inference, the AI service operation policy generates a real-time routing strategy for fast UD data uploading and processing among edge nodes.
For model training, the AI service operation policy determines whether, where, and how to migrate data among edge nodes for achieving data load balancing and improving AI service performance.  

\subsection{Connection with Resource Management}
\label{sec:connection}
Network resources should be properly allocated to support service data management for both inference and model training.
High communication latency in data transmission or high computation latency in processing degrades AI service performance.
Therefore, proper service data management should balance communication and computing loads at each edge node to avoid bottlenecks in data delivery, processing, and training. 
Moreover, service data management consumes communication and computing resources due to data exchange among edge nodes and processing data on edge nodes. Therefore, it is necessary to jointly manage data and resources to support AI service provisioning.




 \begin{figure*}[t]  
  \centering  
  \includegraphics[width=185mm]{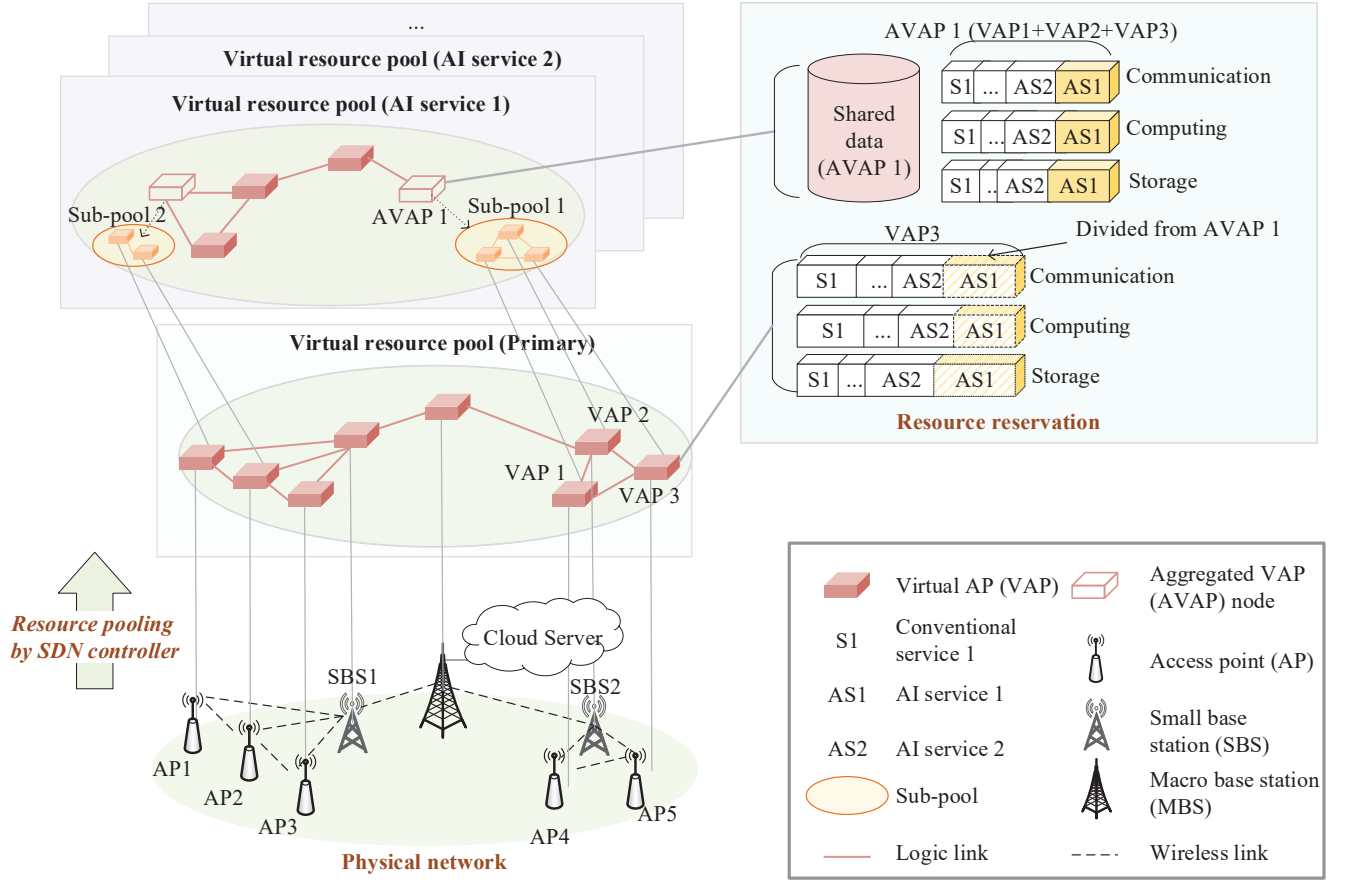}\\
  \caption{Resource pooling and reservation in network slicing.}  
  \label{fig:rv} 
\end{figure*}
\section{Network Slicing for AI Services}
\subsection{Overview}
In network slicing, a software-defined networking (SDN) controller is deployed in the network to create and manage slices for different services and allocates virtual network resources accordingly.
Specifically, network resources are first reserved for slices, referred to as resource reservation, based on service requirements, and subsequently allocated to individual UDs in real time, referred to as resource scheduling~\cite{Zhuang}.  
Although network slicing can support general computing services, further innovations are necessary to support AI services due to their unique features and requirements, as discussed in Section~\ref{sec:AISR}.
In this section, we first discuss the challenges that network slicing faces in supporting AI services. Then, to cope with the challenges, we propose a novel resource pooling method, which is customized for AI services, to refine resource virtualization. Finally, we present a service provisioning approach for AI services by integrating service data management into network slicing.

\subsection{Challenges in AI Service Provisioning}
The location of physical resources affects the performance of AI services at network edge. If computing units in an edge node far away from a UD are selected for inference or training using data from the UD, a long inference latency or a slow learning speed may occur due to the multi-hop communications, which degrades the performance of the corresponding AI service.
In existing network slicing, the location information of physical resources can be hidden in the process of resource abstraction. 
Hiding location information due to abstraction may reduce the capability of network slicing to support edge intelligence.




\begin{figure*}[t]  
  \centering  
  \includegraphics[width=185mm]{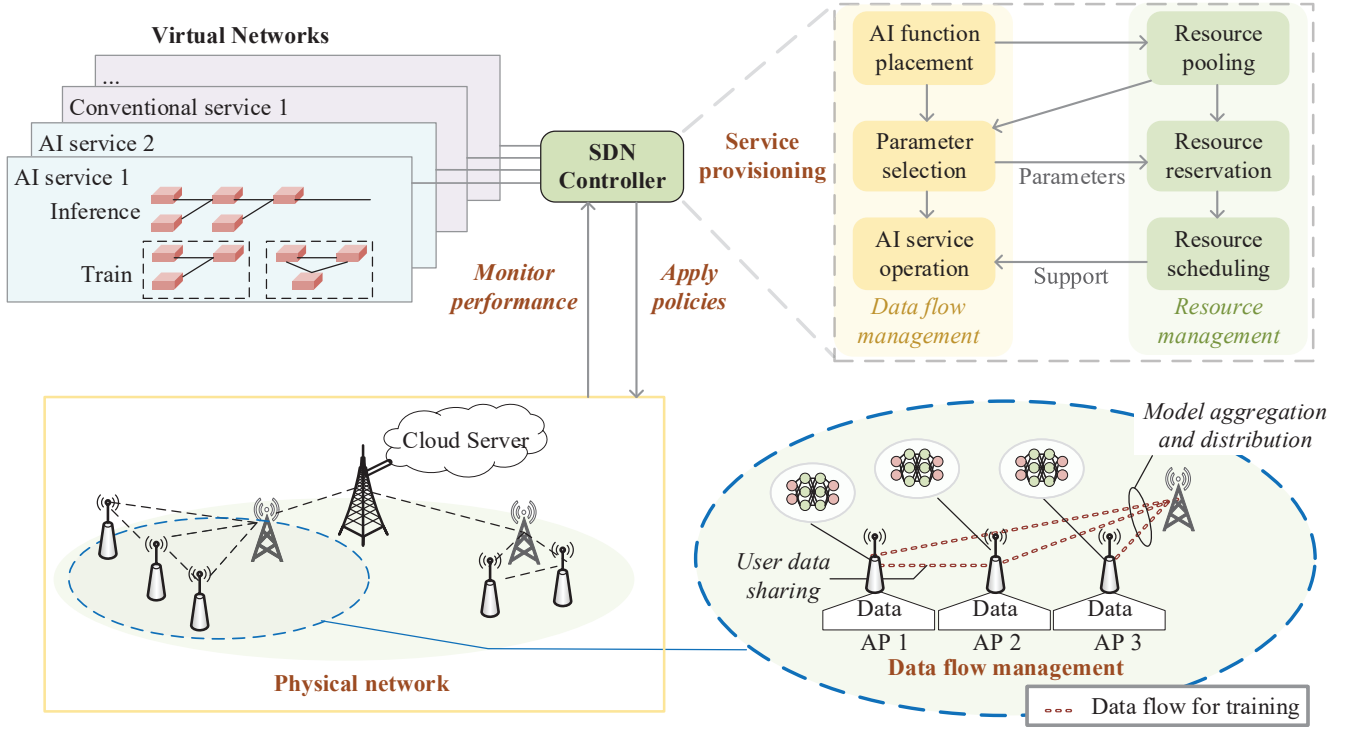}\\
  \caption{Service provisioning for AI services, integrating network slicing and AI service management.}  
  \label{fig:rv2} 
\end{figure*}

As mentioned in subsection~\ref{sec:connection}, data availability at edge nodes impacts the accuracy and learning speed of an AI service, and improving data availability through data migration consumes network resources. Existing network slicing solutions allocate resources without taking service data management into account. Without the coordination of data flow, network slicing cannot satisfy service requirements unique to AI services.

To address these two challenges, we propose a resource pooling method here to refine conventional resource virtualization. The objective is to customize resource virtualization for each AI service according to the location of physical resources and to allocate network resources, while considering data migration among the edge nodes. 


\subsection{Resource Pooling for AI Services}
\label{sec:pooling}


Physical resources in a network can be abstracted to a virtual resource pool via resource pooling, as shown in Fig.~\ref{fig:rv}.
In the virtual resource pool, virtual APs (VAPs) represent the logical servers with computing and storage capabilities and are connected by logical links. The edge nodes, equipped with computing units and storage, are projected to VAPs in the pool.
Virtual network functions (VNFs), as the software implementation of service functions including both AI and conventional functions, are placed at the VAPs. 
A VAP can accommodate multiple VNFs, supported with proper virtual resources for communication, computing, and information storage. 
The resource pool is referred to as primary resource pool.   

Based on the primary resource pool and the physical location of edge nodes, we further abstract physical network resources into customized virtual resource pools, referred to as secondary resource pools for individual AI services. The VAPs in proximity that can support a VNF form a sub-pool to facilitate resource and data sharing for that VNF. The VAPs are aggregated as an aggregated VAP (AVAP) for that VNF in the pool. An example of a secondary resource pool is illustrated in Fig.~\ref{fig:rv}, where sub-pools are formed by VAPs 1 to 3 for a VNF of AI service 1. Within a sub-pool, data collected by VAPs can be migrated among VAPs for inference or model training. 


An AVAP consists of all resources of the VAPs in the corresponding sub-pool. 
During resource reservation, the resources in both the AVAP and VAPs are reserved. Specifically, the resources of an AVAP are first reserved for a VNF to satisfy service requirements. The reservation should account necessary resources for inference, model training, and data migration among VAPs within a sub-pool. Then, VAPs in the sub-pool of the AVAP can flexibly share resources allocated at the AVAP. 
In the example shown in Fig.~\ref{fig:rv}, VAP 1, VAP 2, and VAP 3 are aggregated for AVAP 1 in AI service 1. These VAPs reserve resources for AI service 1 as long as their reserved resources do not exceed the overall resources reserved for AI service 1 allocated at AVAP 1.
While secondary resource pools are used for AI services, conventional services can reserve resources from the primary virtual resource pool. In the above example, the resources at VAP 3 are reserved for all services, including both conventional and AI services. 
During resource scheduling, when the reserved resources in a VAP are not sufficient to support inference or training, data from UDs can be migrated to other VAPs within the same sub-pool for inference or training.

 \begin{figure*}[t]  
  \centering  
  \includegraphics[width=140mm]{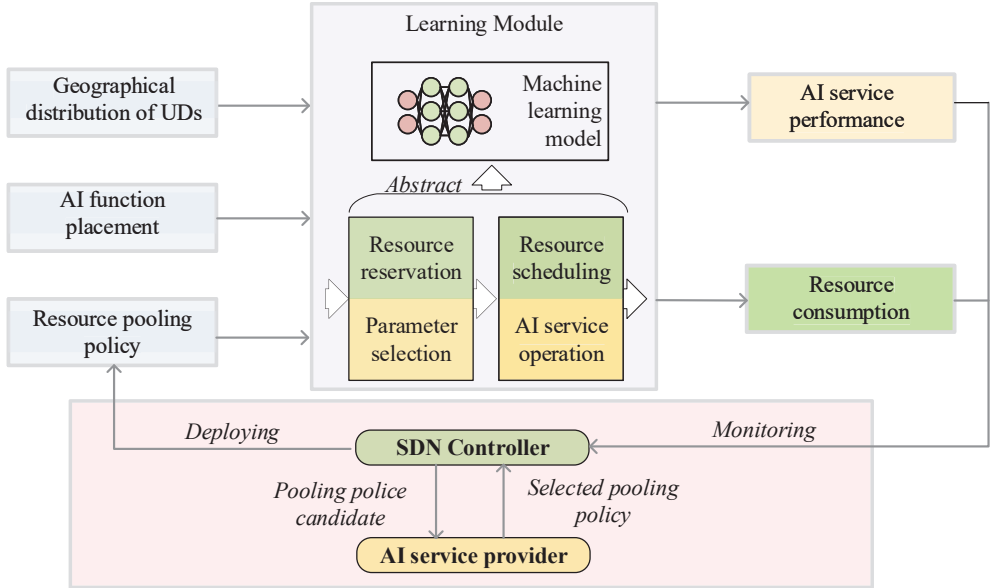}\\
  \caption{Service-oriented resource pool division.}  
  \label{fig:nn} 
\end{figure*}

The main idea of the proposed resource pooling method is to balance the amount of data available to VAPs by data migration among VAPs and to adjust the learning structure of edge intelligence by aggregating resources of VAPs. The goal is to enable service data management in network slicing for satisfying AI service requirements.
The resource pooling policy depends on AI function placement, the geographical distribution of UDs in the network, and the location of the physical resources. 
First, the AI function placement determines which VAPs have the same VNF and thus can be aggregated.
Then, the geographical distribution of UDs and the location of the physical resources determine the amount of data that can be collected by each VAP. Accordingly, the data available to the VAPs can be balanced by migrating data among edge nodes in a sub-pool.
Last, the geographical distribution of UDs and the location of the physical resources affect the network resources consumed for VAP aggregation. 
Specifically, VAP aggregation requires additional communication resources to enable data migration. Meanwhile,  VAP aggregation determines the amount of available data in the corresponding AVAP as well as the computing resources needed for training the data~\cite{cpu1}.

\subsection{Service Provisioning for AI services}

Our AI service provisioning approach combines service data management in subsection~\ref{sec:AIServicemanagement} and resource pooling method in subsection~\ref{sec:pooling}. We illustrate the approach in Fig.~\ref{fig:rv2}. 


An SDN controller is deployed in the network to manage network resources and data flow for AI services.
Firstly, AI functions and corresponding VNFs are placed on edge nodes and corresponding VAPs, respectively. AI function placement policies are adjusted on a large time scale, e.g., days or hours.
Furthermore, according to the physical location of network resources, the geographical distribution of UDs, and AI function placement, secondary virtual resource pools are determined for AI services. AI model parameters are selected according to AI function placement and potential data migration within sub-pools, and resources in the VAP and AVAPs are reserved for different VNFs to meet service requirements. Note that resources for both inference and model training are reserved for the VNFs of AI services.
The resource pooling policy for AI services, AI model parameter selections, and resource reservation are adjusted on a medium time scale, e.g., hours or minutes, to accommodate the spatial-temporal variations of the geographical distribution of UDs.
Last, in real-time network operations, the reserved resources are allocated to individual UDs and network edges according to UD and network dynamics, such as UD mobility and channel conditions. Data from UDs may migrate among VAPs within a sub-pool according to the real-time AI service operation policy, with support from network resource scheduling, to maximize resource utilization and satisfy service requirement. The policies of both resource scheduling and AI service operation are adjusted on a small time scale, e.g., seconds or milliseconds, to allocate resources and manage data flow in real time. 

In the example shown in Fig.~\ref{fig:rv2}, AI functions are deployed at AP 1 to AP 3, where federated learning is adopted for training AI models in the functions. The VAPs, corresponding to AP 1 and AP 2, are in a sub-pool for sharing data collected from UDs. Parameters such as the frequency for model aggregation are determined by the SDN controller, and the network resources are reserved and scheduled correspondingly. Note that AP 1 and AP 2 may train their local models together with data migration for eliminating data bias and improving AI service performance.

\begin{figure*}
        \centering
        \begin{subfigure}[b]{0.495\textwidth}
            \centering
            \includegraphics[width=\textwidth]{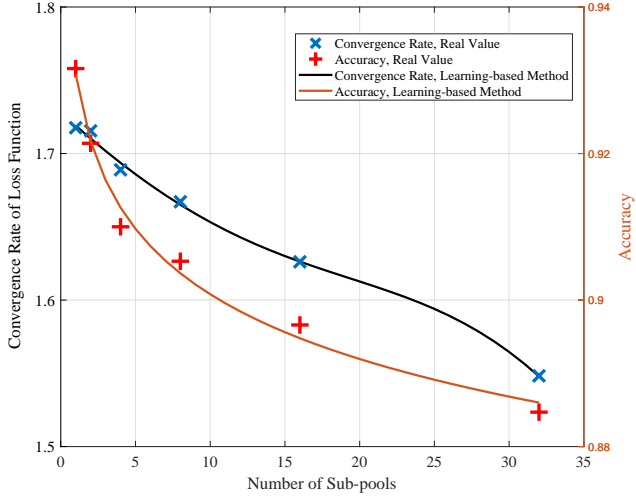}\\
  \caption{Accuracy and the convergence rate of loss of the AI function.}    \label{sim:1a}
        \end{subfigure}
        \hfill
        \begin{subfigure}[b]{0.495\textwidth}  
            \centering 
            \includegraphics[width=\textwidth]{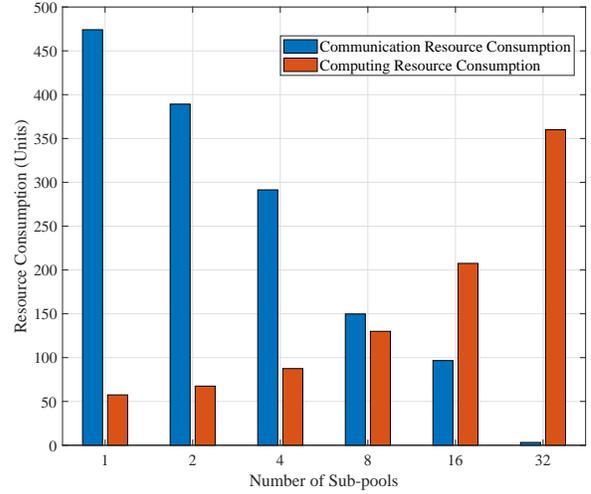}\\
  \caption{Resource utilization for model training in the AI service.}    \label{sim:1b}
        \end{subfigure}
        \caption{Service performance and resource utilization versus the number of sub-pools when $\lambda_{\max}=1$, where accuracy is defined as fraction of correct inferences among the total inferences.}\label{sim:1}
\end{figure*}
\section{Case Study: Service-oriented Resource Pooling}
In this section, we discuss a case study for the proposed resource pooling method. We first present a learning-based method for determining a resource pooling policy. Then, we provide an experiment to demonstrate the effectiveness of resource pooling policies.
\subsection{Learning-based Resource Pooling}
\label{sec:cs}
As mentioned in Section~\ref{sec:pooling}, network resources are reserved and scheduled for AI services from secondary resource pools. With different resource pooling policies, secondary resource pools and the resulting AI service performance and resources consumption will also differ. 
Therefore, as illustrated in Fig.~\ref{fig:nn}, we utilize a learning module, supported by machine learning techniques, e.g., DNNs, to learn the relation between resource pooling policy and the corresponding AI service performance and resource consumption, given the resource allocation and service data management strategies.
The inputs of the learning module include AI function placement policy, resource pooling policy for all AI services, and the geographical distribution of UDs during a time interval between two successive resource pooling policy updates. 
The outputs are the performance and the average resource consumption of the AI services during the time interval.
The learning module is trained at the SDN controller. Specifically, the SDN controller deploys different resource pooling policies, monitors corresponding AI service performance and resource consumption, and uses the monitored information to further train the learning module.
When the learning module is fully trained, the SDN controller selects resource pooling policy candidates that yield satisfactory AI service requirements with minimum resource consumption. Then, an AI service provider chooses a resource pooling policy from the candidates, and the SDN controller deploys the selected policy in the network. The criteria for selecting the resource pooling policy can be service-specific. For example, privacy-sensitive AI functions may choose a pooling policy that reduces data migration as much as possible.

\subsection{Numerical Results}
\subsubsection{Experiment Setups}
We conduct trace-driven simulations to evaluate the proposed resource pooling method and the corresponding learning-based resource pooling policy determination. In the considered network, there are 32 APs on network edge. Each AP has deployed the same AI function for inference. The AI functions are trained using federated learning. Specifically, APs gather data from UDs and train their local models once every second. A macro BS gathers the parameters of local models from APs once every ten seconds, generates a global model using the FedAvg algorithm~\cite{li2019convergence}, and distributes the parameters of the global model to all APs. The content of the AI function used in the simulation is handwritten-digit recognition with dataset from the MNIST database~\cite{dataset}. The AI model in a function includes 3 fully connected layers with 784, 200, and 10 neurons, respectively. The learning rate for the local model is 0.01, and the optimizer is stochastic gradient descent. In our simulation, the data collected by different APs is non-i.i.d. Moreover, to learn the resource pooling policies, we use the Gaussian process regression method as the learning module mentioned in subsection \ref{sec:cs}.

We use different aggregated arrival rates of data for UDs at different APs.  The data arrival rate at an AP is randomly selected from (0,$\lambda_{\max}$], where $\lambda_{\max}$ denotes the maximum data arrival rate. Each AP corresponds to a VAP in the primary resource pool. 
We change the number of sub-pools in the secondary resource pool to adjust the pooling policy. Based on the number of sub-pools, the VAPs are grouped by the \textit{k}-means method to form sub-pools according to the physical locations of the APs and the data arrival rates at the APs.
Resource requirements are summarized as follows: one resource unit (RU) is consumed for transmitting one unit of data between any two APs; 0.5 RU is consumed for processing one unit data for training; 0.1 RU is consumed for offloading and distributing DNN models in federated learning. Moreover, 10 RUs are consumed for training a DNN model. The SDN controller reserves resources accordingly based on the average data arrival rates and schedules the resources. During resource scheduling, additional cost is applied if the reserved resources become insufficient in real time. 

 
\subsection{Performance Evaluation}

The impact of resource pooling policy on the AI service performance and resource consumption for training is shown in Figs.~\ref{sim:1a} and~\ref{sim:1b}, respectively. As the number of sub-pools decreases, model training requires more communication resources but less computing resources for training. This is because more APs migrate their collected data, which generates additional cost on communication, while fewer APs train their local models, which reduces the overall computing resource consumption.
Moreover, by aggregating data into fewer APs, the model training is conducted in a more centralized learning structure, and data bias can be eliminated by load balancing. Therefore, a higher training speed and a higher accuracy can be achieved under a pooling policy with a lower number of sub-pools. Moreover, we utilize a learning module to model the relation between resource pooling policy and AI service performance, as presented in subsection~\ref{sec:cs}, and the performance approximated by the learning module is accurate, as shown in Fig.~\ref{sim:1a}. 

 \begin{figure}[t]  
  \centering  
  \includegraphics[width=88mm]{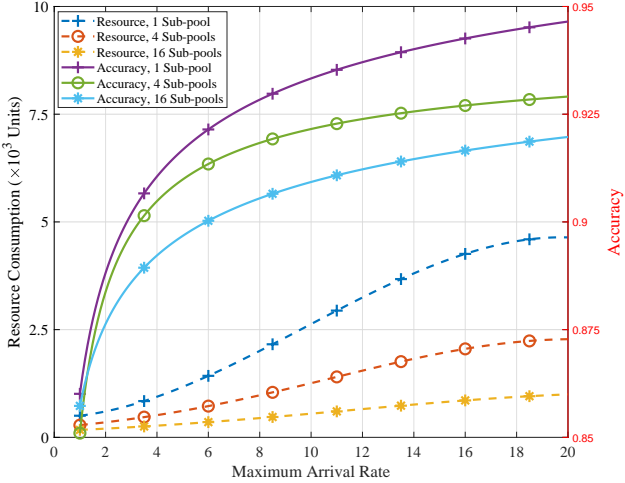}\\
  \caption{Service performance and average resource consumption versus user data arrival rates.}  
  \label{sim:2} 
\end{figure}

The AI service performance and average resource consumption with different $\lambda_{\max}$ and resource pooling policies are shown in Fig.~\ref{sim:2}. As $\lambda_{\max}$ increases, the resource consumption increases due to the need for processing more data in training. Meanwhile, with a lower arrival rate, the accuracy of the AI service degrades. This is because the available data for training at each AP decreases, and overfitting happens when a small amount data is trained with a high learning rate. 

\section{Conclusion}

In this article, we investigate AI service provisioning on network edge for 6G.  AI services depend on data for training and inference, and, because of such dependence, AI service provisioning requires joint management of data and conventional network resources. Accordingly, within the framework of network slicing, we propose a resource pooling method to connect data and network resources in AI service provisioning. The proposed method first supports data management in network slicing while balancing between AI service performance and resource consumption of data management. In addition, the proposed method considers location of physical resources in resource virtualization for network slicing. With our approach, the network and the service providers can determine where and how to train AI models depending on the data availability and resource constraints in the network as well as the service performance requirements.



{\footnotesize
\bibliographystyle{IEEEbib}
\bibliography{reference}
}

\end{document}